\pdfoutput=1
\documentclass{emulateapj}
\usepackage{graphics,graphicx}
\usepackage{fancyheadings}
\usepackage{color}

\usepackage{natbib}
\usepackage{amsmath}
\usepackage{hyperref,breakurl}
\usepackage{epsfig}

\shorttitle{Thermal plasma in Centaurus~A}
\shortauthors{O'Sullivan et al.}

\begin{document}

\title{Thermal plasma in the giant lobes of the radio galaxy Centaurus~A}

\author{S.~P.~O'Sullivan$^{1,2\ast}$, I.~J.~Feain$^{1,2}$, N.~M.~McClure-Griffiths$^{1}$, R.~D.~Ekers$^{1}$, E.~Carretti$^{1}$, T.~Robishaw$^{3}$, S.~A.~Mao$^{4,5}$, B.~M.~Gaensler$^{2}$, J.~Bland-Hawthorn$^{2}$, \L.~Stawarz$^{6,7}$ }

\medskip

\affil{$^{1}$CSIRO Astronomy and Space Science, ATNF, PO Box 76, Epping, NSW 1710, Australia}
\affil{$^{2}$Sydney Institute for Astronomy, School of Physics, The University of Sydney, NSW 2006, Australia}
\affil{$^{3}$Herzberg Institute of Astrophysics, Dominion Radio Astrophysical Observatory, PO Box 248, Penticton, BC V2A 6J9, Canada}
\affil{$^{4}$Jansky Fellow, National Radio Astronomy Observatory, P.O. Box O, Socorro, NM 87801, USA}
\affil{$^{5}$Department of Astronomy, University of Wisconsin, Madison, WI 53706, USA}
\affil{$^{6}$Institute of Space and Astronautical Science JAXA, 3-1-1 Yoshinodai, Chuo-ku, Sagamihara, Kanagawa 252-5210, Japan}
\affil{$^{7}$Astronomical Observatory, Jagiellonian University, ul. Orla 171, 30-244 Krak\'ow, Poland}

\medskip

\email{s.o'sullivan@physics.usyd.edu.au.}
\label{first page}

\begin{abstract}

We present a Faraday rotation measure (RM) study of the diffuse, polarized, radio emission 
from the giant lobes of the nearest radio galaxy, Centaurus A. 
After removal of the smooth Galactic foreground RM component, 
using an ensemble of background source RMs located 
outside the giant lobes, we are left with a residual RM signal associated with the giant lobes.  
We find the most likely origin of this residual RM is from thermal material mixed throughout 
the relativistic lobe plasma. 
The alternative possibility of a thin-skin/boundary layer of magnetoionic material swept 
up by the expansion of the lobes is highly unlikely since it requires, at least, an 
order of magnitude enhancement of the swept up gas over the expected 
intragroup density on these scales. 
Strong depolarisation observed from 2.3 to 0.96~GHz also supports the presence 
of a significant amount of thermal gas within the lobes; although depolarisation 
solely due to RM fluctuations in a foreground Faraday screen on scales smaller 
than the beam cannot be ruled out. 
Considering the internal Faraday rotation scenario, we find a thermal 
gas number density of $\sim10^{-4}$~cm$^{-3}$ implying 
a total gas mass of $\sim10^{10}$~M$_\odot$ within the lobes. The thermal pressure 
associated with this gas (with temperature $kT\sim0.5$~keV, obtained from 
recent X-ray results) is approximately equal to the non-thermal pressure, 
indicating that over the volume of the lobes, there is approximate equipartition 
between the thermal gas, radio-emitting electrons and magnetic field (and potentially 
any relativistic protons present). 

\end{abstract}


\section{Introduction}
The integrated history of a radio galaxy is encoded in the magnetized, relativistic gas 
within its lobes. These lobes have been inflated by outflows emanating from the 
central supermassive black hole (SMBH) of the host galaxy \citep{begelmanblandfordrees1984} 
with their structure illuminated by the synchrotron emission we detect in 
the radio band. 
Through their formation, radio galaxies can deposit $10^{55}$--$10^{62}$~erg of mechanical energy into  
the intergalactic medium (IGM) over the lifetime of the source \citep{mcnamara2009}, 
as well as potentially enriching the IGM with heavy elements \citep{aguirre2001,reuland2007} 
and magnetic fields \citep{furlanettoloeb2001}. 
Determining exactly how radio galaxies transfer energy and material to the IGM is essential for 
understanding their impact on both galaxy and cosmic structure evolution \citep{bower2006, best2006, croton2006}. 
In particular, knowledge of the fraction of ``non''-radiating particles (thermal gas and relativistic protons)
is of primary importance for understanding the pressure balance 
and dynamics of radio galaxy lobes with respect 
to their environments \citep{morganti1988, dunn2005, birzan2008, croston2008}, 
as well as their potential to provide seed particles for high energy emission 
and ultra-high energy cosmic rays \citep{fermicena}. 

Faraday rotation provides a 
sensitive diagnostic for the presence of magnetized, ionized thermal material. 
Previous Faraday rotation studies 
\citep{perley1984, spangler1985, laingbridle1987, garringtonconway1991, kronberg2004, feain2009} 
have placed upper limits on the uniform number density of thermal gas within radio galaxy lobes of $\lesssim10^{-4}$~cm$^{-3}$, 
assuming the minimum-pressure magnetic field intensity. 
However, conclusively disentangling the different contributions from Faraday rotation 
that is internal and/or external to radio galaxy lobes has proven very difficult. 

In this paper, we present complementary spectropolarimetric observations of the nearest radio galaxy Centaurus A 
at 1.4~GHz with both the Parkes 64~m single-dish radio telescope and the Australia Telescope Compact Array (ATCA) 
aperture synthesis telescope.  
The Parkes observations 
are sensitive to diffuse emission which allows us to measure the 
Faraday rotation measure (RM) of the polarized emission from Centaurus~A itself, while the high angular resolution 
ATCA observations \citep{feain2009} provide us with an 
ensemble of RMs from 281 background radio sources that are 
uncontaminated by any diffuse polarized emission and are located along 
sightlines both inside and outside the giant lobes. 
Centaurus~A is the only radio galaxy for which such an analysis 
is possible with current sensitivity due to its large angular size 
of $\sim$$9^{\circ}\times4^{\circ}$. 
For a distance to Centaurus~A of $\sim$3.8~Mpc \citep{harris2010}, $1^{\circ}$ corresponds 
to a projected linear size of $\sim$66~kpc. 

In Section 2, we describe the observations, calibration and RM analysis. 
Section 3 outlines our results along with a discussion on various 
models and the implication for the giant lobes. 
Our conclusions are presented in Section 4.

\section{Observations and Data Reduction}

\subsection{Parkes Data}
A $10^{\circ}\times14^{\circ}$ area centred on the position 
of Centaurus~A was observed using the Parkes 64~m telescope at 1.4~GHz over a period 
of 10 days (2009 March 2--11), totalling 80 hrs of telescope time. 
The Parkes data used in \cite{feain2011} were archival but the new 
observations presented here were required because the archival data 
did not contain polarisation information. 
The observations were conducted using a cross-scanning technique 
with the H-OH receiver and a quarter-wave plate, similar to \cite{mao2012}, 
providing 0.25~MHz channels across 256~MHz of bandwidth centerd on 1388~MHz. 
The total useable bandwidth covered 1312 to 1480~MHz. 
The FWHM for the Parkes telescope at 1380~MHz is approximately $14.4'$. Data were 
obtained while scanning the telescope at a rate of $3^\circ.5$~min$^{-1}$ and sampling 
every 1~s, meaning that data were recorded every $7'$. This resulted in 120 right ascension 
(RA) scans and 86 declination (Dec) scans. The data were then imaged with $8'\times8'$ pixels 
to ensure at least one data sample per pixel (some blank pixels occurred towards the Northern 
edge of the image where the scanning rate was slightly too fast for $7'\times7'$ pixels).  
This effectively smooths the image and we measure an actual gridded beam width 
of $\sim16'$ at 1312~MHz.

Calibration of all scans was done 
using the Parkes Continuum Polarimetry Software 
(\textsc{ParkesPol})\footnote{https://svn.atnf.csiro.au/trac/parkespol} 
package. The source PKS~B$1934-638$ was used for flux and bandpass 
calibration due to its bright, stable and well known flux level of 14.95~Jy at 1380 MHz \citep{reynolds1994}. 
Basic flagging was done on all scans before the 0.25~MHz channels 
were rebinned into 8~MHz channels and the polarization calibration 
was performed on each channel separately. Ten scans of 3C\,138 
at a range of feed angles from $-45^{\circ}$ to $+45^{\circ}$ 
provided a sufficient range of parallactic angle coverage 
to calibrate the instrumental polarizations. 
The polarisation angle calibration was 
achieved using the known polarization angle of PKS B$0043-424$ 
($+143.3^{\circ}$ at 1.4~GHz, ${\rm RM}=+2$~rad~m$^{-2}$, 
Carretti \& Haverkorn, private communication). 

Stokes $I$, $Q$ and $U$ images 
were created for each frequency 
channel by combining all RA and Dec scans. Initially, separate RA 
and Dec maps were created with a linear baseline removal performed 
using the edge of the image. This removes both the ground emission 
contamination as well as any structure on scales the size of the map and 
greater. The final maps are produced by combining the orthogonal scan maps 
in Fourier space using the technique of \cite{emersongraeve1988} which 
minimizes the effects of baseline drifts on raster scanned data. 

\subsection{ATCA Data}
\cite{feain2009} observed an $\sim$45~deg$^2$ area centred on the host galaxy 
NGC~5128 with the Australia Telescope Compact Array (ATCA) at 1.4~GHz. 
They derived RMs for 281 background radio sources using essentially the 
same spectropolarimetric techniques as employed in this paper (see Section~2.3) with 
$24\times8$~MHz channels covering 1288 to 1480 MHz. 
In order to analyse the background radio sources without contamination 
from the large scale emission from Centaurus~A, they filtered out the emission 
on large spatial scales by disregarding all visibilities from baselines shorter than 
300~metres (1.4~k$\lambda$).
The high angular resolution ATCA observations ($\sim$$40''$) thus provided 
an ensemble of background radio sources that were uncontaminated by 
any diffuse polarized emission and located along sight-lines both 
inside and outside the edge of the giant lobes.

\begin{figure}
    \includegraphics[width=8.0cm]{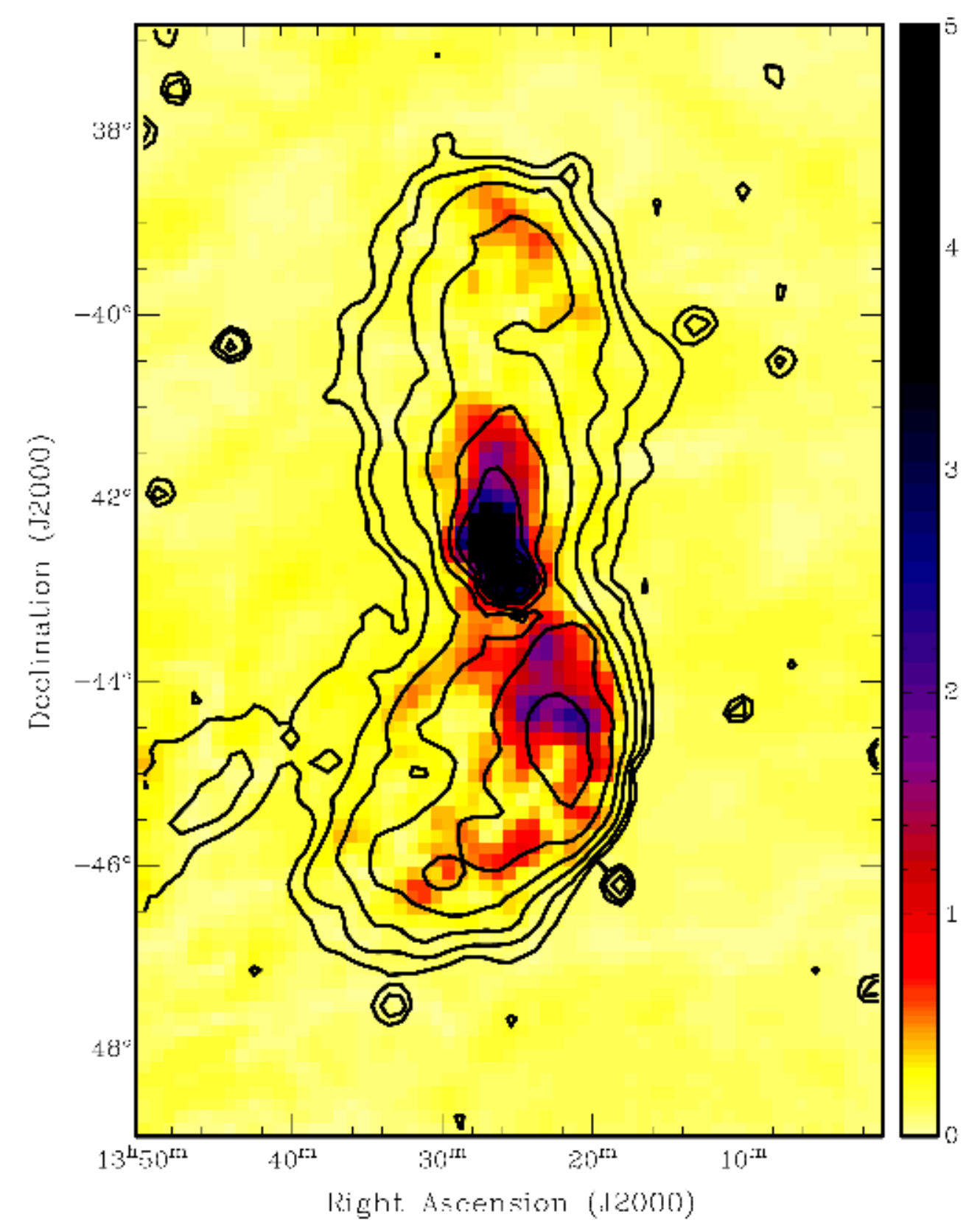}
  \caption{Polarized intensity (in Jy~beam$^{-1}$) of the full observed region 
overlaid with the 1.4~GHz total intensity contours starting at 250~mJy~beam$^{-1}$ 
and increasing by factors of two. Diffuse polarized emission is detected throughout the 
entire region due to the Galactic foreground. 
The degree of polarization of the strongly polarized regions 
($p>300$~mJy~beam$^{-1}$) within the lobes ranges from $\sim$10--40\%.  
}
  \label{fig_p}
\end{figure}

\begin{figure}
    \includegraphics[width=8.0cm]{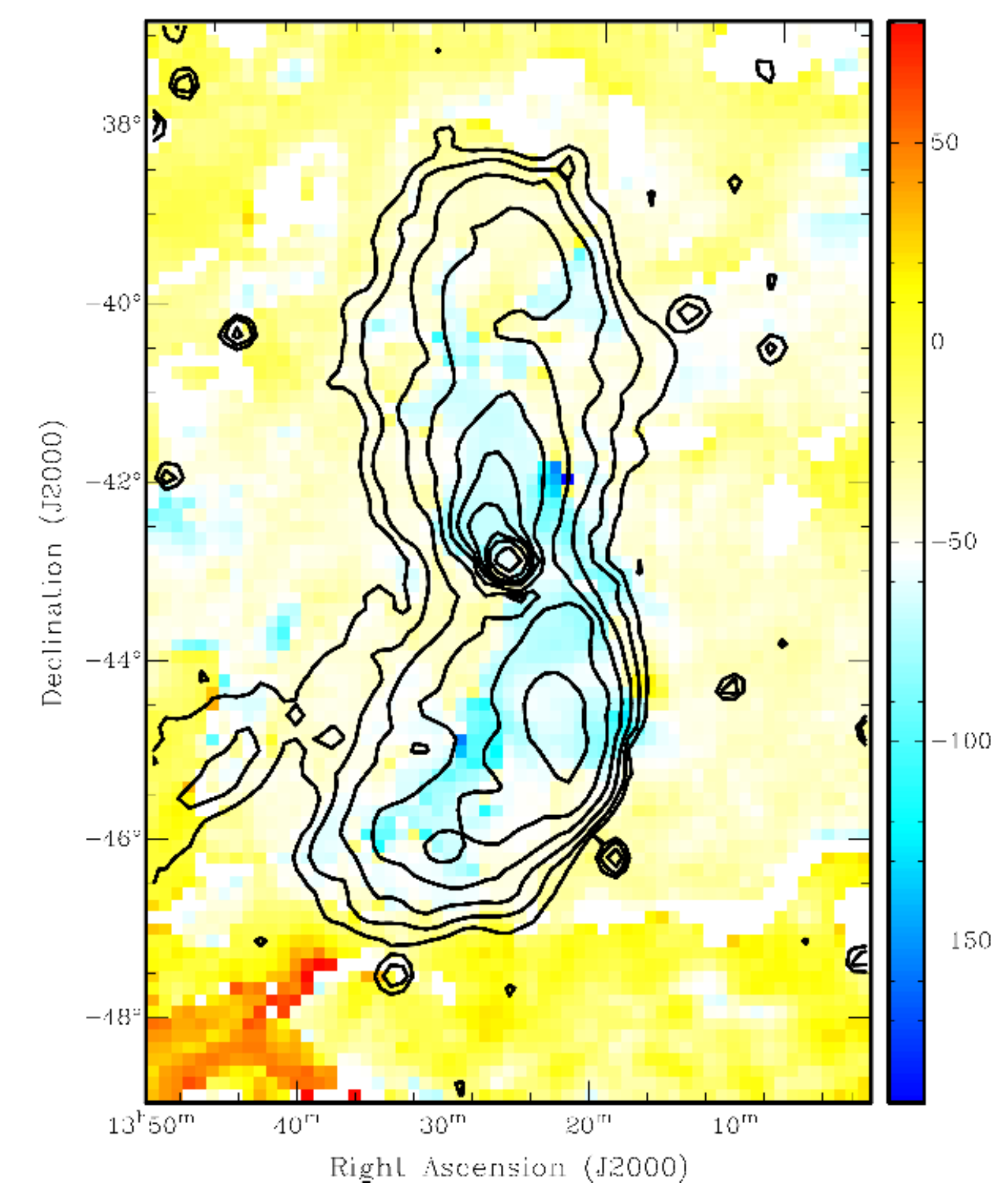}
\caption{Faraday rotation measure (RM) derived at each $8'\times8'$ pixel, 
overlaid with the 1.4~GHz total intensity contours. 
The range of RM on the color scale goes from $-175$ to $+100$~rad~m$^{-2}$. 
}
\label{fig_RM}
\end{figure}

\subsection{Rotation Measure Analysis}
 
Faraday rotation provides a direct diagnostic 
for the presence of magnetized thermal material
though observations of the change in the state of polarization 
with wavelength as the radiation passes through magnetoionic media on 
its path to us. The Faraday depth ($\phi$) of a particular region of polarized 
emission is defined as 

\begin{equation} \label{FaradayDepth}
\phi(L) = 0.81 \int^{0}_{L}{ n_e {B_{||}} dl} ~~{\rm rad~m}^{-2},
\end{equation}

\noindent where $n_e$ is the electron number density in cm$^{-3}$, $B_{||}$ is the line-of-sight 
magnetic field strength in $\mu$G and $L$ is the distance through the magnetoionic region in parsecs. 
The total observed Faraday rotation measure (RM), 
defined as $d\chi/d\lambda^2$, where $\chi$ is the polarisation angle, 
can have contributions from multiple 
regions along the line of sight. 
In the simplest case of a uniform external Faraday screen, the RM and 
Faraday depth are identical and the change
in the polarization angle follows the relation 
\begin{equation}
\chi=\chi_0+{\phi}\,\lambda^2,
\end{equation}
\noindent where $\chi_0$ is the intrinsic ($\lambda=0$) polarization angle. 

To determine the RM we first constructed cubes of Stokes $Q$ and $U$ 
vs.~$\lambda^2$ and applied the RM synthesis and RMCLEAN \citep{bdb2005, heald2009} 
techniques following the algorithm described by \cite{osullivan2012}. 
This generated a cube of polarized intensity ($p=\sqrt{Q^2+U^2}$) at all 
Faraday depths from $-3000$ to $+3000$~rad~m$^{-2}$ from which 
the RM image was generated by extracting the RM at the peak 
polarized intensity. 
With the correct RM at each pixel the cube of polarized intensity is 
collapsed to form the polarized intensity image with the full sensitivity of the 168~MHz band. 

Figure~\ref{fig_p} shows the diffuse polarized emission detected from Centaurus~A as 
well as from the Galaxy, which fills the entire field. 
The resolution in Faraday depth space ($\delta\phi$) of our experiment, defined by 
the FWHM of the Rotation Measure Spread Function (RMSF), is 310~rad~m$^{-2}$. 
The largest magnitude Faraday depth we can reliably detect is $\sim$3000~rad~m$^{-2}$ 
and the maximum detectable Faraday thickness is $\sim$80~rad~m$^{-2}$. 

For a typical signal-to-noise ratio ($S/N$) of 50 to 100 on source, the error in 
the RM estimate at each pixel is $\sim$1--3~rad~m$^{-2}$ 
($\Delta RM\sim \delta \phi/(2\times S/N)$). 
This error estimate relies on the assumption that there is 
only one dominant Faraday depth component within the FWHM of the 
RMSF; if there are multiple Faraday depth components within our RMSF 
then we cannot uniquely specify their locations to better than 310~rad~m$^{-2}$. 
However, for large $S/N$ ratios the RMCLEAN algorithm may in principle be 
able to resolve components within the RMSF. We do not find any evidence 
for such components or for any significant broadening of the RMSF within 
the lobes of Centaurus~A. 
In some regions of the observed field we find more than one peak in our 
Faraday depth spectrum. While the origin of these multiple Faraday depth components 
along the line of sight is interesting, none of these regions lie within the lobes 
of Centaurus~A and are not included in our current study. 

\section{Results \& Discussion}

The measured RMs from the Parkes observations are shown in Figure~\ref{fig_RM} displaying 
a clear enhancement in the magnitude of the RM within the lobes of Centaurus~A. 
Note that even with no Faraday rotating material in the vicinity of Centaurus~A, we 
could observe such an enhancement in the magnitude of the observed RM in regions 
where the polarized emission from Centaurus~A dominates over the Galactic emission. 
This can occur because the 
polarized emission of Centaurus~A probes the full Faraday depth along the line of sight 
to us, while the diffuse Galactic polarized emission originates somewhere within our 
Galaxy and may only probe more local Faraday rotating regions depending 
on the exact location(s) of the polarized emission within our Galaxy. 
However, due to potential changes in the line of sight magnetic field direction, 
a longer line of sight does not necessarily mean a larger observed Faraday 
depth. 
The Galactic emission 
may also be subject to strong internal Faraday rotation effects resulting in a significantly 
lower RM than that measured from the polarized emission from Centaurus~A. 

\subsection{Residual RM from Centaurus~A}

\begin{figure}
    \includegraphics[width=8.5cm]{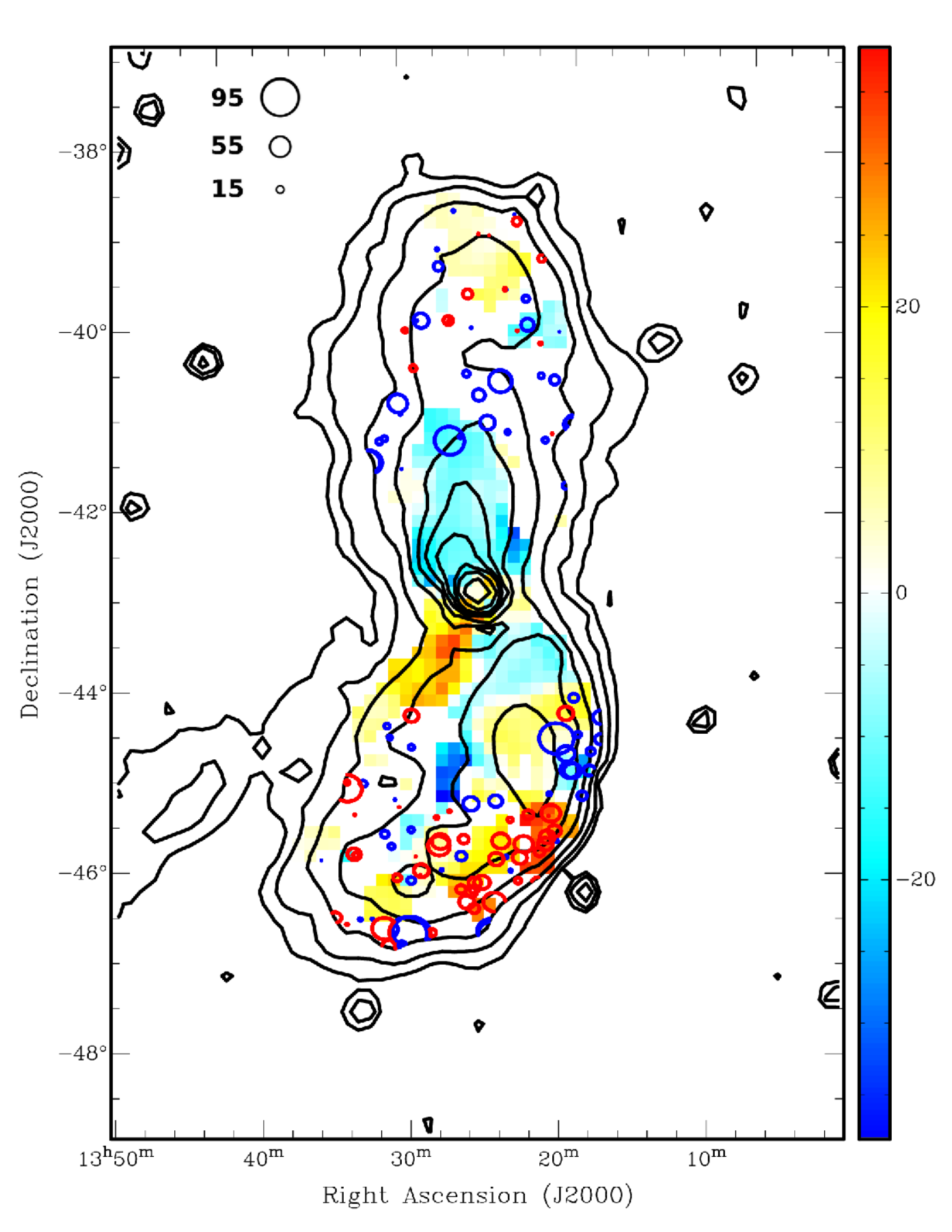}
\caption{The residual RM (color scale ranging from $-40$ to $+40$~rad~m$^{-2}$) 
associated with Centaurus~A after subtraction of the Faraday rotation 
  due to foreground magnetoionic material. 
   The residual RMs of the background sources inside the lobes are represented by circles 
 whose size corresponds to the magnitude of the RM while red/blue circles indicate 
 positive/negative signed RMs. The legend in the top left corner gives the magnitude 
 of the RM, in rad~m$^{-2}$, in relation to the size of the circles. 
 The are no point source RMs towards the inner regions of the lobes because 
 these regions were masked out due to imaging errors mainly 
 caused by the bright core; see \cite{feain2009} for details. 
}
\label{fig_RRM}
\end{figure}

The mean RM of the background sources outside the lobes is $-52.9$~rad~m$^{-2}$ 
with a standard deviation of $29.2$~rad~m$^{-2}$ \citep{feain2009}. Therefore, the 
Faraday depth of our Galaxy is the dominant contributor to the total observed RM. 
The RMs from the 160 background sources located outside the lobes were used to 
remove the smooth part of the foreground Galactic Faraday rotating material. 
This was achieved by fitting a first-order two-dimensional polynomial to the point-source 
RMs outside the lobes. 
By subtracting this RM surface, with a gradient of $\sim6$~rad~m$^{-2}$~deg$^{-1}$, 
from both the background source RMs (inside and outside the lobes) 
and the RM of Centaurus~A, we obtain the residual RM signal. 
The colored pixels in Figure~\ref{fig_RRM} show the residual RM signal from the 
polarized emission of Centaurus~A while the open circles show the residual 
RM from the background sources inside the lobes. 

The mean of the magnitude of the residual RM from the polarized 
emission from Centaurus~A greater than 300~mJy~beam$^{-1}$ is 
$\langle |{\rm RM}| \rangle=12.0\pm0.3$~rad~m$^{-2}$ 
(using the standard error of the mean)\footnote{Using only the same 
sightlines as the background sources we 
get $\langle |{\rm RM}| \rangle=13.0\pm1.9$~rad~m$^{-2}$.}. 
The mean value of the RM from the Galactic polarized emission 
outside the lobes is approximately $-27$~rad~m$^{-2}$ with a 
standard deviation of 23~rad~m$^{-2}$ 
while the mean polarized intensity is $\sim52$~mJy with a standard 
deviation of 20~mJy. 
As the Galactic polarized emission becomes a significant 
fraction of the Centaurus~A polarized emission towards the edges 
of the lobes, the systematic error in the measured RM will 
become large. 
Since we are unable to resolve the two contributions 
with our current dataset, we have only included in our analysis 
the regions within the lobes where the polarized emission from 
Centaurus~A is strong enough to make the effect from the Galactic 
emission negligible (i.e., for Centaurus~A polarized emission 
$\geq$300~mJy~beam$^{-1}$, see Appendix).  

It is possible that the residual RM signal could be explained by a degree-scale 
variation in the Galactic Faraday depth that happens to coincide with the orientation 
of the giant lobes and is not sampled by the background sources outside the lobes. 
We consider this unlikely but cannot definitively rule out such a scenario. 
There is a clear asymmetry in the distribution of signs of the residual RM 
between the Northern and Southern lobes. 
However, we are cautious not to over-interpret the sign of the residual RM since 
the smooth, foreground RM surface subtraction will not have 
removed any small angular-scale variations in the RM of our 
Galaxy. Therefore, residual RM contamination from our Galaxy may 
contribute to local enhancements in the residual RM magnitude as well as 
changes in the sign of the residual RM in patches throughout the lobes. 
Therefore, we defer a detailed analysis of individual features of RM 
amplitude and sign to a more detailed study of the 
polarisation properties of the giant lobes.

\begin{figure}
    \includegraphics[width=8.5cm]{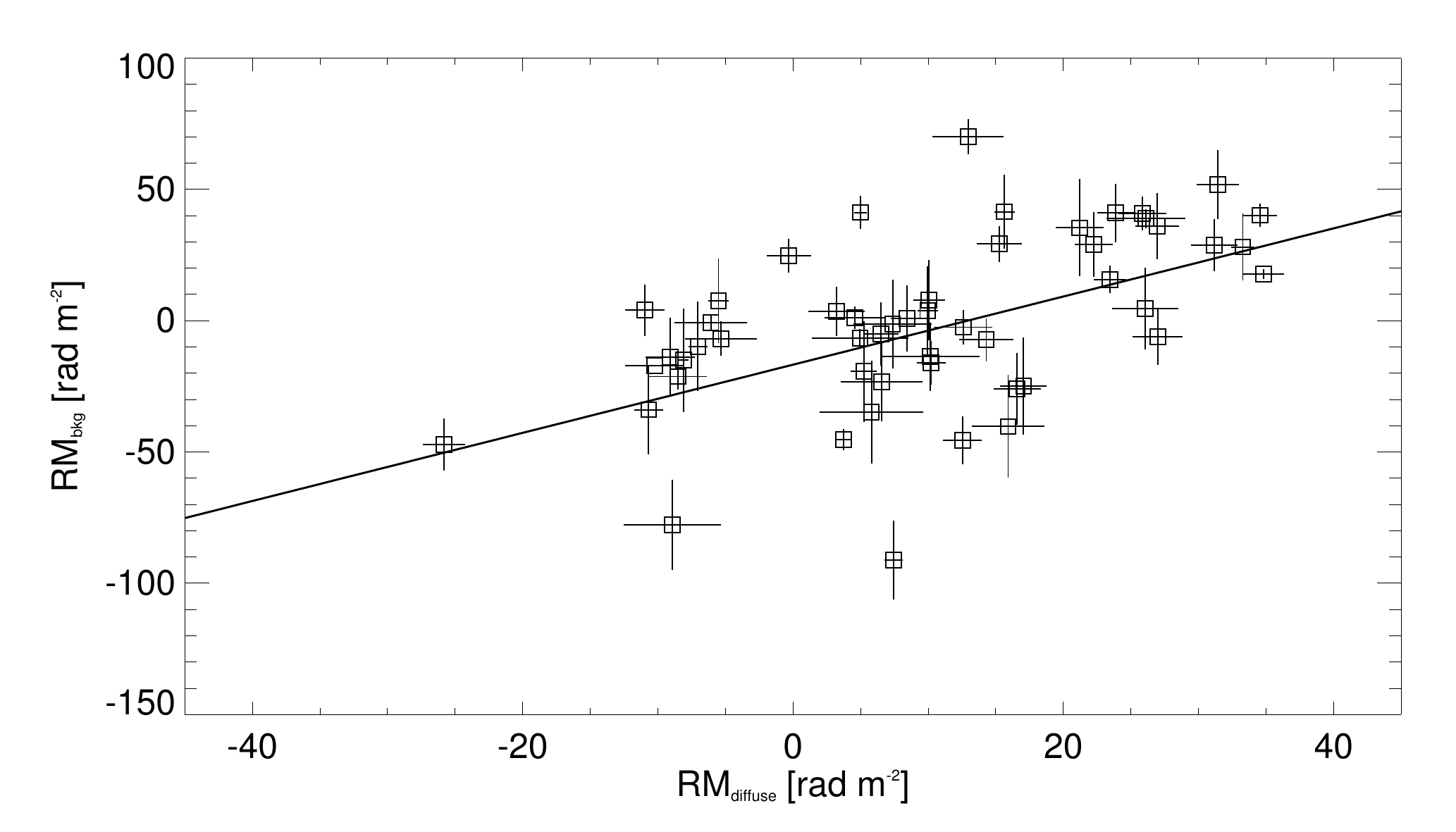}
\caption{Plot of the background source residual RMs (RM$_{\rm bkg}$) 
versus the residual RMs from the polarized emission of Centaurus~A (RM$_{\rm diffuse}$) 
at the same position, with 
both axes in rad~m$^{-2}$. Solid line represents a best-fit line with slope $1.3\pm0.1$. 
}
\label{fig_correlation}
\end{figure}

\begin{figure}
    \includegraphics[width=8.0cm]{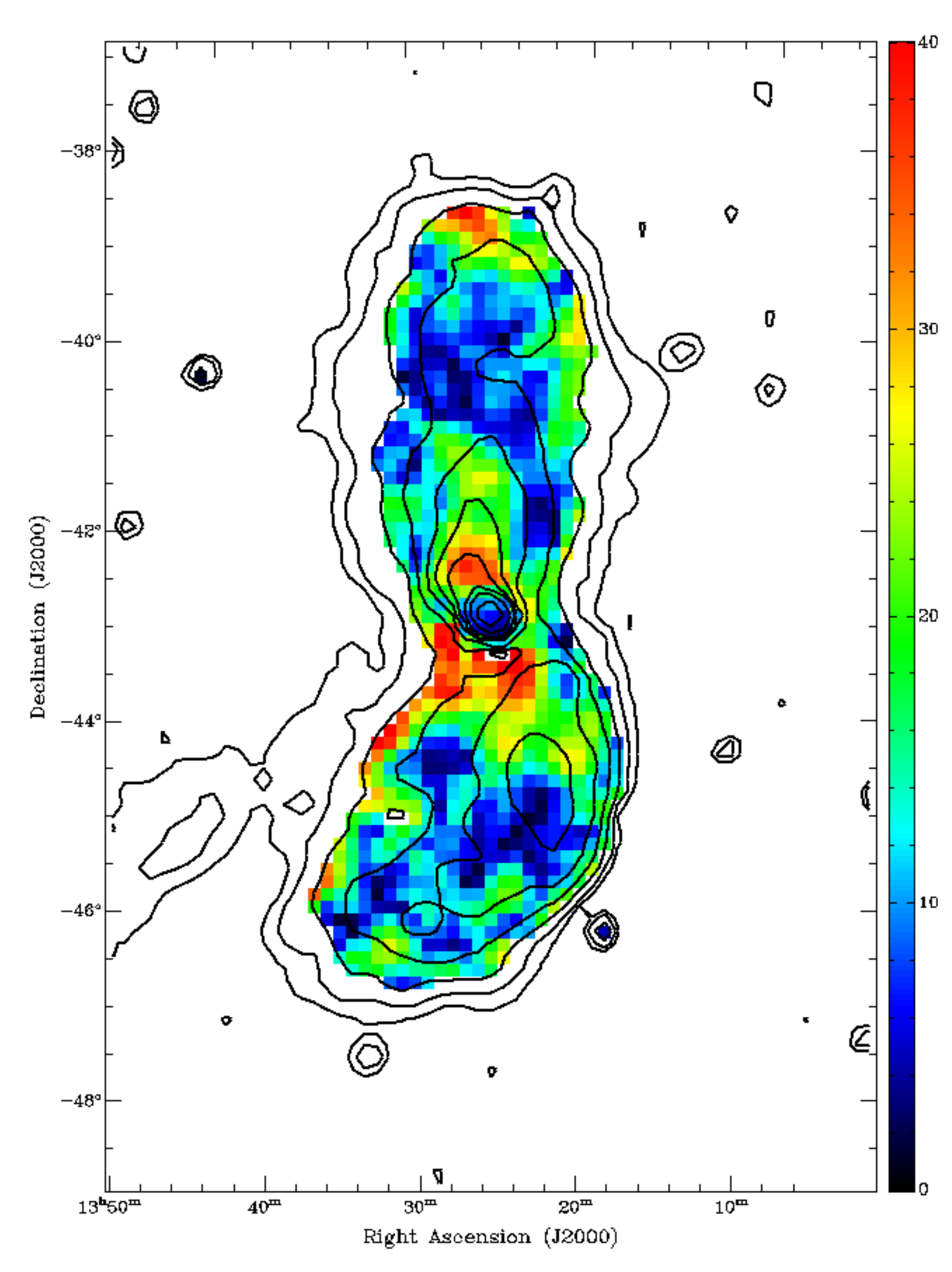}
\caption{
 Fractional polarisation image ranging from 0--40\% 
 overlaid by the total intensity contours. 
 A particularly interesting feature is the high 
 fractional polarisation located $\sim$0.5$^\circ$ 
 South-West of the core. 
 The fractional polarisation is not shown out to 
 the edge of the total intensity contours because 
 a dramatic increase that is unrelated to the physical 
 properties of Centaurus~A is observed. 
 The effect is as a result of the steep falloff in total intensity while 
 the Galactic polarized emission remains approximately constant. 
}
\label{fig_fpol}
\end{figure}

\subsection{Origin of the Residual RM Signal}

The observed residual RM signal may be due to Faraday rotation 
from thermal gas within the lobes, a thin boundary layer of swept 
up material surrounding the lobes or a large-scale fluctuation in the 
Galactic foreground Faraday depth coinciding with the position of 
the giant lobes. 
It is important to note that the subtraction using the fit to the background 
RMs just outside the lobes not only removes the smooth Faraday depth contribution 
of our Galaxy but also of any smooth RM component on large angular scales caused by 
the intragroup medium or an extended halo of magnetized thermal plasma surrounding the lobes. 

The sign of the residual RM of the background sources within the lobes 
appears to correlate well with the RM from the diffuse emission (Figure~\ref{fig_RRM}). 
To quantitatively investigate the relation, we plot the residual background source 
RMs (RM$_{\rm bkg}$) against the corresponding residual RMs of the 
polarized emission from Centaurus~A (RM$_{\rm diffuse}$) along the same sightline (Figure~\ref{fig_correlation}). 
There are some obvious discrepancies, which are not unexpected 
given the different scales probed by the ATCA ($\sim$$50''$) and Parkes ($\sim$$16'$) 
observations as well as the intrinsic scatter in the background source RMs \citep{feain2009}. 
A correlation-coefficient of 0.54 between these two quantities shows that 
both the sign and amplitude of the two different RM measurements are related, 
with a best fit to the data giving ${\rm RM_{bkg}}\sim1.3 {\rm RM_{diffuse}}$. 

Considering only the lines of sight in which the sign of RM$_{\rm bkg}$ and 
RM$_{\rm diffuse}$ agree, we find a median ratio, 
$q\equiv {\rm RM_{bkg}}/{\rm RM_{diffuse}} \sim 1.5$. 
The median, rather than the mean, is quoted to avoid the 
results being skewed by a small number of outliers. 
Such a relation, with $q > 1$, is expected for most internal Faraday 
rotation models with $q=2$ in the case of uniformly mixed emitting and 
rotating regions \citep[e.g.,][]{cioffi1980}. 
However, a value of $q\sim2$ is also expected in the case of a thin 
boundary layer of magnetoionic material surrounding the lobes. 
This is assuming that the lobes and their immediate environments 
are symmetric in the sense that any postulated 
Faraday rotating region on the near side of the lobe is replicated on the far 
side. 
For the Northern lobe only, $q\sim1.3$, while for the Southern lobe $q\sim1.5$. 
This tentatively suggests that either a larger amount of gas is present within the 
Southern lobe or the lobe orientation is such that the Southern lobe is further away; 
however, it is difficult to claim this difference as significant 
given the smaller number of data points in the Northern lobe 
compared to the Southern lobe.  
Overall, the range $1.3\leq q\leq1.5$ supports an interpretation 
of internal Faraday rotation in which there is an asymmetric 
distribution of emitting and rotating regions within the lobes \citep[e.g.,][]{sokoloff1998}, 
possibly with a small amount of Faraday rotation external to the lobes. 
However, we cannot discount entirely external effects due to the orientation of the 
lobes or residual RM contamination from the Galaxy (as described in Section 3.1). 
We avoid any in-depth modelling based on the value of $q$ due to the 
sparse and unevenly sampled nature of the background sources. 
A much denser grid of background sources is required for more robust 
comparisons and detailed modelling of the distribution of emitting and rotating 
regions associated with the giant lobes.

Previous work by \cite{feain2009} found a turbulent RM signal associated 
with the southern lobe due to either turbulent structure 
throughout the lobe or in a thin skin surrounding the lobe boundary. 
If we assume that the residual large-scale RM signal we detect 
is due to a thin boundary layer of depth $\delta\sim20$~kpc, then with 
an estimate of the line-of-sight magnetic field, $B_{||}$, we 
can obtain the expected electron number density of the skin, $n_{e,skin}$. 
Based on the value of the magnetic field strength in the lobes of $B\sim0.9$~$\mu$G, 
as derived in \cite{fermicena} based on broadband modelling of radio and $\gamma$-ray 
data, we can place a strong upper limit for the line-of-sight magnetic field in the skin 
of $B_{||} \sim B/\sqrt{3} \sim 0.5$~$\mu$G. 
This leads to a conservative lower limit of $n_{e,skin} \sim 1.5\times10^{-3}~({\rm RM}/12~{\rm rad~m^{-2}})(B_{eq}/0.5~\mu{\rm G})^{-1}(\delta/20~{\rm kpc})^{-1} $~cm$^{-3}$. 
This type of model is mainly based on the work of \cite{bicknell1990} who 
used hydrodynamic simulations to show that Kelvin-Helmholtz 
instabilities may form on the surface of radio lobes and potentially 
cause advection of the lobe magnetic field into the surrounding medium. 

The lobes of Centaurus~A extend into a sparse group of galaxies \citep{russianguy} with 
the Centaurus~A host galaxy, NGC~5128 at the centre. 
An intergalactic medium (IGM) density of $n_{igm}\sim10^{-3}$~cm$^{-3}$ was suggested by 
\cite{bouchard2007} to explain a discontinuity in the HI properties of 
Centaurus A group dwarf galaxies through ram-pressure stripping arguments. However, this 
would seem excessively large for such a poor group of galaxies.  
In order to estimate a more plausible intragroup density at the extent 
($r$) of the giant lobes, we use a standard profile $n_{igm} (r ) \sim n_0 (r/a_0)^{-b}$.  
This gives $n_{igm}(200~{\rm kpc})\sim 10^{-4}$~cm$^{-3}$ 
for the typically derived values of $n_0\sim10^{-2}$~cm$^{-3}$, $a_0\sim10$~kpc and $b\sim1.5$ \citep{mulchaey1998, sun2012}, 
although the large scatter in the scaling parameters should be noted. 
The expansion of the lobes is expected to sweep up and possibly 
compress the intragroup medium around the edges of the lobes. 
Even if we consider a maximum compression factor of four, from a strong 
shock (which would be highly unlikely for the giant lobes), it 
is still not sufficient to reach $n_{e,skin}\sim10^{-3}$~cm$^{-3}$. 
Furthermore, a rather low magnetisation of the IGM is expected on scales 
of hundreds of kiloparsecs from the group centre, certainly 
much less than the equipartition value in the 
lobes, unless some additional processes are able to substantially 
amplify the IGM magnetic field, as discussed in \cite{bicknell1990} for 
example. 
Many theoretical models for the structure of extended 
lobes actually assume zero magnetisation of the ambient medium into which 
the lobes evolve \citep[e.g.,][]{gourgouliatos2010}. Hence, the estimated 
value of $n_{e,skin}\sim10^{-3}$~cm$^{-3}$ should be considered as 
a very conservative lower limit, corresponding to the maximum magnetisation 
of the IGM in the vicinity of the expanding lobes. 
Therefore, from the above arguments we conclude that the observed 
residual RM signal cannot be entirely explained by a thin skin scenario 
and that a significant fraction of thermal gas must be mixed throughout 
the lobe volume.

\begin{figure}[!hb]
    \includegraphics[width=8.5cm]{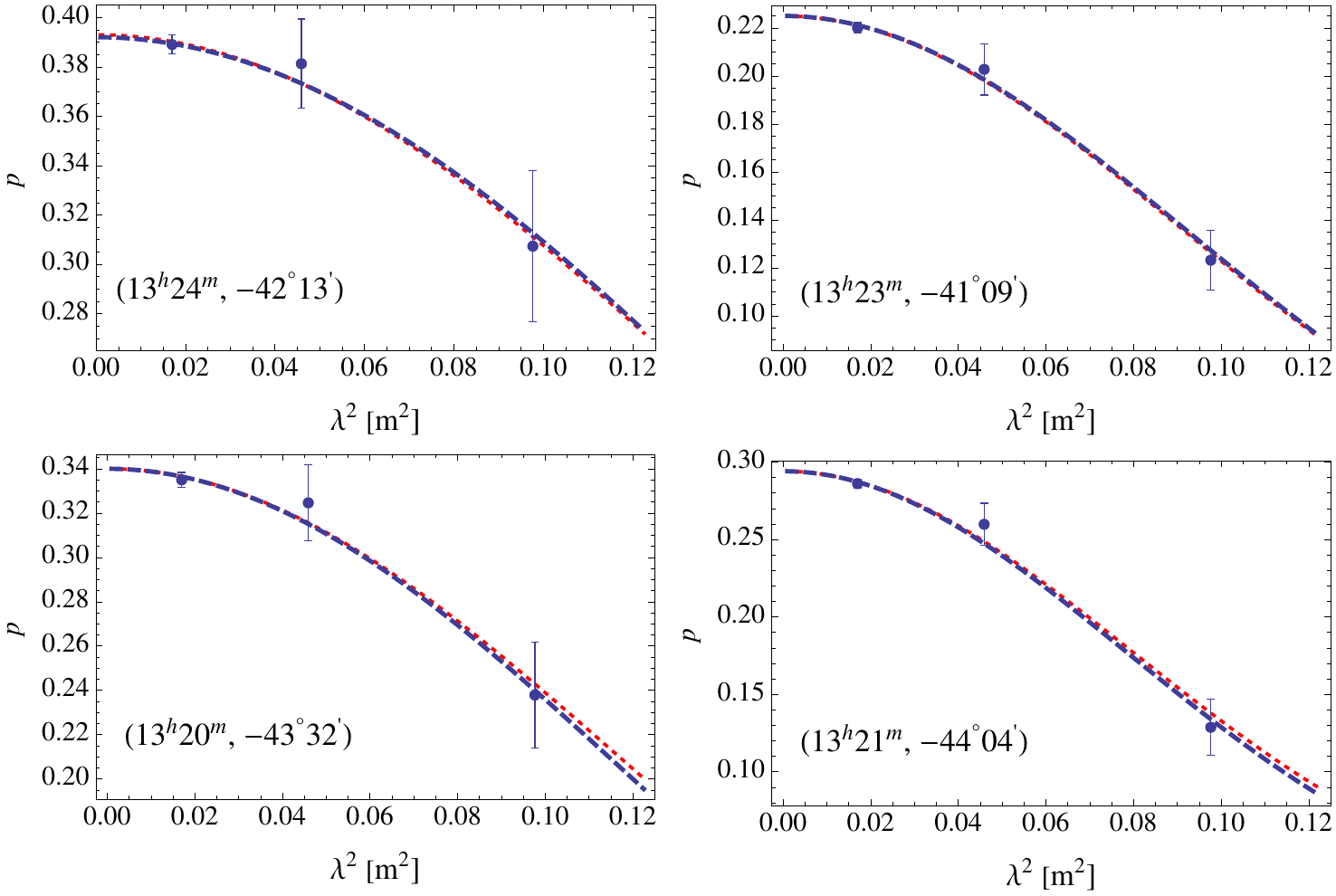}
\caption{Plots of the degree of polarisation ($p$) versus wavelength 
squared ($\lambda^2$) for four regions of the giant lobes of Centaurus~A 
at 2.3, 1.4 and 0.96~GHz.  
We chose areas with high S/N in $p$ as well as small variations in 
RM across an area equivalent to the 0.96~GHz beamsize. 
The dotted (red) curves represent a model of depolarisation solely from 
external Faraday dispersion while the dashed (blue) curves represent a 
model combining both internal Faraday rotation and external 
Faraday dispersion. In the top right and top left plots, the dashed (blue) 
and dotted (red) curves are almost identical. 
}
\label{fig_depol}
\end{figure}

\subsection{Depolarization of the Lobes}
From the residual RM signal alone, we cannot determine exactly how much 
thermal material is likely to be mixed in with the relativistic lobe plasma. 
A potential discriminant comes from analysing the change in fractional 
polarisation with wavelength. 
For mixed radio-emitting and Faraday-rotating regions, the polarization 
angle of radiation produced at different depths within the lobes is rotated 
by different amounts which leads to wavelength-dependent depolarisation 
\citep{burn1966, sokoloff1998}.  
A thin-skin or boundary layer effectively acts as a foreground Faraday screen 
and does not depolarize the emission from the lobe \citep{bicknell1990}. 
An unfortunate complication for the foreground screens is an effect known as 
external Faraday dispersion \citep{burn1966, tribble1991}. This occurs 
when many turbulent cells of magnetoionic material are within the telescope 
beam resulting in different amounts of Faraday rotation along different lines of 
sight which causes depolarisation when averaged across the beam area.  

In order to better constrain the origin of the residual RM signal, 
we analysed the change in the degree of polarisation between 
2.3, 1.4 and 0.96~GHz. The 2.3~GHz image was obtained from 
the S-band Polarization All Sky Survey (S-PASS), which recovers 
the absolute level of the polarized emission across the Southern sky 
\citep{carretti2011} 
and hence provides a very robust estimate for the level of polarized 
emission from Centaurus~A at 2.3~GHz. 
The 0.96~GHz fractional polarisation image was created using 
the Stokes $I$ and $p$ contour plots from \cite{cpc1965}. 
Digital copies of these images are not available so we used the 
algorithm of \cite{westphalen1995} to reliably represent the 
contour plots in digital form. Given that the noise is not recovered 
in the digitized images, we limit our analysis to the highest 
signal-to-noise (S/N) areas where the estimated error is $\lesssim10\%$ \citep{westphalen1995}. 
Both the 2.3 and 1.4~GHz images were smoothed, in $Q$ and $U$ 
to the resolution of the 0.96~GHz image.

\subsection{External Depolarisation Model}

For our depolarisation analysis we have chosen four positions 
which have high S/N as well as small observed variation in 
RM across the beam so that the depolarising effects of 
external Faraday dispersion are minimized.  
To model the depolarising effect of a foreground screen 
with a Gaussian random magnetic field we use 
\begin{equation}
p=p_0 \exp(-2\sigma_{\rm RM}^2 \lambda^4), 
\end{equation}
where $p_0$ is the intrinsic degree of polarized emission 
from the lobes and $\sigma_{\rm RM}$ describes the RM 
fluctuations on scales smaller than the observing beam. 

Figure~\ref{fig_depol} shows the fractional polarisation versus wavelength 
squared for two positions in the Northern lobe, ($13^h27^m$, $-42^{\circ}29'$)
and ($13^h26^m$, $-41^{\circ}25'$), and two in the Southern lobe, 
($13^h23^m$, $-43^{\circ}49'$) and ($13^h24^m$, $-44^{\circ}21'$). 
For the chosen positions, the variation in RM across an area the 
size of the 0.96~GHz beam is less than the estimated error of 
the individual RM measurements. 
The dotted (red) curves in Figure~\ref{fig_depol} represent a fit to the data 
using equation~3 with values of $\sigma_{\rm RM}$ ranging from 
3.5--6.3~rad~m$^{-2}$. Table~1 lists the values of the fitted parameters 
with their respective error estimates. 
From this we see that depolarisation from an external Faraday 
screen can, in principle, explain the total depolarisation 
observed at these four positions of the giant lobes. 
However, we note that fluctuations in the complex polarisation can become 
dominant at long wavelengths resulting in a much weaker than 
expected decrease in the degree of polarisation due to external 
Faraday dispersion \citep{tribble1991, sokoloff1998}. Hence, 
for Centaurus~A, the observed depolarisation may not be very well 
described by an exponential decrease in the degree of polarisation. 

In Figure~\ref{fig_p} we can see a strong asymmetry between the polarized 
intensity in the inner parts of the Northern and Southern giant lobes, the 
origin of which could be attributed to the 
Laing-Garrington effect \citep{laing1988,garrington1988}. 
This effect is commonly assigned to a halo of thermal gas, 
threaded by a turbulent magnetic field, surrounding the 
radio galaxy \citep{garringtonconway1991}. In this case, the depolarisation 
is larger for the assumed more distant Southern lobe, in which 
the path length through the surrounding gas is larger. 
However, there is no compelling evidence for the 
exact orientation of the lobes. 
The effect may also be explained by internal differences 
between the two lobes and/or from a gradient, with Galactic latitude, 
in the properties of the foreground Galactic Faraday screen. 

In order to comment further on this issue, we show the fractional polarisation 
within the giant lobes in Figure~\ref{fig_fpol}. One of the most interesting 
features of this image is the high degree of polarisation located $\sim$0.5$^\circ$ 
South-West of the bright core. 
We speculate that this region may be the oppositely-directed counterpart of the 
well studied Northern middle lobe \citep{morganti1999}, which until now 
had no observed Southern counterpart.
This ``Southern middle lobe'' may have lost its structure and merged with 
the outer/older radio lobe after 
uplifting and mixing with a large amount of gas from the host galaxy. 
This scenario would require some asymmetry in the distribution of gas within the host 
galaxy which is not unreasonable in the aftermath of a merger event \citep{sparke1996}. 

In fact, a recent study by \cite{bellcomeau2012} found, for sources 
exhibiting the Laing-Garrington effect, that the jet brightness asymmetry 
on kpc-scales cannot be explained by beaming and therefore must 
be intrinsic (in cases where the estimated kpc-scale outflow speeds 
are close to $0.1c$). 
They conclude that a Laing-Garrington effect 
due to intrinsic differences in the Faraday rotating material 
within radio galaxy lobes cannot be discounted. 
Based on these results and our arguments in Section~3.2 for a significant fraction of 
thermal material mixed in with the relativistic plasma of the lobes, 
we next consider a model in which a significant fraction 
of the observed depolarisation (and residual RM signal) is 
due to internal Faraday rotation. 


\begin{deluxetable}{ccc}
\tablecaption{\label{table_EFD} External Faraday dispersion model parameters.}
\tabletypesize{\scriptsize}
\tablewidth{0pt}
\tablehead{\colhead{Position} & \colhead{$p_0$} & \colhead{$\sigma_{\rm RM}$} \\  
                                                       & (\%) & (rad~m$^{-2}$)}
\startdata
($13^h27^m$, $-42^{\circ}29'$) & $39.3\pm0.4$ & $3.5\pm0.3$ \\
($13^h26^m$, $-41^{\circ}25'$) & $22.5\pm0.2$ & $5.5\pm0.3$ \\
($13^h23^m$, $-43^{\circ}49'$) & $34.0\pm0.5$ & $4.2\pm0.4$ \\
($13^h24^m$, $-44^{\circ}21'$) & $29.4\pm0.6$ & $6.3\pm0.5$ 
\enddata
\tablecomments{Col.~1: Position of extracted data in RA and Dec in J2000. 
Col.~2: intrinsic degree of polarisation. 
Col.~3: variation in the external RM on scales smaller than the beam. }
\end{deluxetable}

\begin{deluxetable}{cccc}
\tablecaption{\label{table_IFR} Internal Faraday rotation model parameters.}
\tabletypesize{\scriptsize}
\tablewidth{0pt}
\tablehead{\colhead{Position} & \colhead{$|\phi|$} & \colhead{$\sigma_{\phi}$} & \colhead{$\sigma_{\rm RM}$} \\ 
     				          	& (\%) & (rad~m$^{-2}$)    & (rad~m$^{-2}$) }
\startdata
($13^h27^m$, $-42^{\circ}29'$) & $10.2\pm1.7$ & $1.7\pm0.7$ & $1.2\pm0.6$ \\
($13^h26^m$, $-41^{\circ}25'$) & $10.0\pm0.7$ & $3.0\pm0.2$ & $4.1\pm0.3$ \\
($13^h23^m$, $-43^{\circ}49'$) & $10.1\pm1.2$ & $2.1\pm0.4$ & $2.7\pm0.4$ \\
($13^h24^m$, $-44^{\circ}21'$) & $10.1\pm0.9$ & $2.1\pm0.6$ & $5.5\pm0.4$ 
\enddata
\tablecomments{Col.~1: Position of extracted data in RA and Dec in J2000. 
Col.~2: internal Faraday depth of the lobes. 
Col.~3: internal Faraday dispersion within the lobes. 
Col.~3: variation in the external RM on scales smaller than the beam.}
\end{deluxetable}

\subsection{Internal Depolarisation Model}

If the observed Faraday rotation and depolarisation are due to thermal 
gas within the lobes, we can estimate the number density of the gas using 
an internal Faraday rotation model that accounts for 
depolarisation from the ordered magnetic field ($B_{ord}$) 
as well as from the random magnetic field ($B_{rdm}$) 
within the lobes. In this model, we consider the simplest case of a Gaussian 
distribution of Faraday depths with mean $\phi$ and 
standard deviation $\sigma_\phi$ in a slab with a linear depth $L$  
along the line of sight \citep{burn1966, sokoloff1998}. 
The intrinsic ($\lambda=0$) fraction of polarized emission from Centaurus~A 
($p_0$) is then modified following

\begin{equation}
p=p_0\left( \frac{1-e^{-2\sigma_\phi^2 \lambda^4 +2i\lambda^2 \phi}}{2\sigma_\phi^2\lambda^4-2i\lambda^2\phi} \right)e^{-2\sigma_{\rm RM}^2 \lambda^4},
\end{equation}

\noindent where $\sigma_\phi=0.81n_eB_{\rm rdm}(Ld)^{1/2}$ with $d$ 
representing the scale of magnetic field fluctuations of the random 
magnetic field ($B_{\rm rdm}$) such that the 
number of cells along the line of sight is $L/d$.  We have also included 
the effect of external Faraday dispersion, as described in Section~3.4, 
and use the same values of $p_0$. 

The dashed (blue) curves in Figure~\ref{fig_depol} represent fits to the 
data using the above model for each of the four lobe positions. 
In all cases, we find the magnitude of the internal Faraday depth, 
$|\phi|\sim10$~rad~m$^{-2}$ with $\sigma_\phi$ 
ranging from 1.7 to 3.0~rad~m$^{-2}$ and $\sigma_{\rm RM}$ ranging 
from 1.2 to 5.5~rad~m$^{-2}$. 
Since we have only three data points and have to fit for three variables, 
this essentially guarantees a good fit, perhaps giving false 
confidence in the appropriateness of the model.
Hence, we consider the fitted parameters of this model, shown in Table~2,  
as reasonable estimates of the model 
parameters which can describe the observed depolarisation 
given our current observational knowledge. 

Due to the random component of the magnetic field in the 
lobes, we expect to see an excess dispersion in the RM within the lobes 
over the RM dispersion outside the lobes (i.e., $\sigma_\phi$). However, 
this is difficult to disentangle from the turbulent magnetoionic medium 
of our Galaxy that likely varies throughout the entire field as well 
as from effects more local to the source due to the orientation of the 
lobes (e.g. the Laing-Garrington effect). 
The standard deviation of the RM from the diffuse Galactic emission 
(outside the lobes) is $\sim$$12$~rad~m$^{-2}$ while the RM of the polarized 
emission from Centaurus~A has a standard deviation of $\sim$$14$~rad~m$^{-2}$. 
This difference cannot be regarded as significant, given our individual measurement 
errors (Section~2.3), but it is not inconsistent with the fitted values of $\sigma_{\phi}$ 
used to describe the depolarisation in the lobes. 

The total magnetic field strength ($B_{tot}$) in the Northern and Southern 
lobes has been estimated at $\sim0.9$~$\mu$G from the modelling 
of radio and $\gamma$-ray data \citep{fermicena}. 
The intrinsic degree of polarization of synchrotron radiation ($p_{i}$) is 
reduced from its maximum level 
according to the relation $p_0=p_{i} B_{ord}^2/B_{tot}^2$ 
where $B_{tot}^2=B_{ord}^2+B_{rdm}^2$ and 
$p_{i}=(3-3\alpha)/(5-3\alpha)$, with $\alpha$ representing the spectral index \citep{burn1966}. 
Therefore, using the spectral index of particular regions of the giant lobes obtained 
from \cite{hardcastle2009} and the mean observed fractional polarization 
from our observations, we can estimate the strength of the 
ordered ($B_{ord}$) and random ($B_{rdm}$) magnetic field components. 
For example, at the position ($13^h23^m$, $-41^{\circ}09'$) we have $p_0=22.5\%$, therefore 
$B_{ord}\sim B_{rdm}\sim B_{||}\sim0.5$~$\mu$G. 
Considering a path length through the lobes of $L\sim200$~kpc and a 
magnetic field fluctuation scale of $d\sim20$~kpc, 
we use $\sigma_\phi=(5\times10^4) n_e B_{\rm rdm}$ and 
$\phi\sim (2.5\times10^5) n_e B_{||}$ to find, for the four selected regions 
of the lobes, $n_e\sim (0.9-1.3)\times10^{-4}$~cm$^{-3}$. 

\cite{feain2009} provided an upper limit on the volume-averaged 
thermal electron density of the lobes of 
$\left<n_e\right>\lesssim 1.3\times10^{-4} (0.5/B_{||})^{-1}$~cm$^{-3}$, 
where in this case we use $B_{||}\sim0.5$~$\mu$G instead of the 
equipartition value of 1.3~$\mu$G used in their paper. 
This limit was based on sampling of the background sources inside the lobes, probing 
a distribution presumably more inhomogeneous than the sparse sampling of the 
background sources could detect. However, the limit is not inconsistent with 
our results. 
Our measured value of $n_e\sim10^{-4}$~cm$^{-3}$ based on our internal 
Faraday rotation interpretation provides a very good description of all the 
current observables (i.e., the residual RM signal, the depolarisation 
and the excess RM dispersion associated with the lobes). 
Although we cannot definitively rule out external Faraday effects 
as an explanation for the total amount of observed depolarisation, 
we prefer the internal Faraday rotation interpretation since it better 
explains all the observational evidence. Taking this as our preferred 
model we now investigate some of the important consequences for 
the giant lobes. 

\subsubsection{Implications for the Giant Lobes}

In order to calculate the pressure in the lobes due to the thermal gas 
($p_{\rm th}\sim n_ekT$), we require an estimate for the temperature 
($T$) of the gas. Recent X-ray results from a small region of the southern lobe found 
an excess of diffuse, thermal emission from Centaurus~A over the background, 
with a best-fit gas temperature of $kT\sim0.5$~keV 
and a number density of $\sim10^{-4}$~cm$^{-3}$ \citep{stawarz2012}, in excellent 
agreement with the density estimate made in Section~3.3 above.  
Interestingly, this leads to a thermal pressure, $p_{th}\sim8\times10^{-14}$~erg~cm$^{-3}$ 
which is approximately equal to the total non-thermal pressure, 
$(p_{e\pm}+U_B)\sim8\times10^{-14}$~erg~cm$^{-3}$ \citep{fermicena, stawarz2012}, 
where the magnetic energy density $U_B\equiv B_{tot}^2/8\pi$. 
It is important to note that our derivations do not depend on an equipartition 
assumption, and therefore, on the unknown contribution of relativistic protons. 
They depend instead on the magnetic field value derived from the broadband modelling in 
\cite{fermicena}, whose primary assumption is that the detected $\gamma$-ray flux is solely 
due to the inverse-Compton emission of the radio emitting electrons. 
The derived magnetic field value is thus consistent with the case of an approximate energy 
equipartition between the magnetic field, radiating electrons, and potentially 
any relativistic protons present (i.e. $U_B \sim U_e \gtrsim U_p$). 

The majority of extended lobes in low-power radio galaxies are found to be under-pressured 
or in approximate pressure equilibrium with their external environment, under the assumption 
of energy equipartition between the magnetic field and radio-emitting electrons within the 
lobes \citep{morganti1988, croston2008}. 
Hence, those lobes must 
be either far from the magnetic field--relativistic electrons energy equipartition, or there must 
be additional pressure provided by particles contributing only weakly to the observed 
non-thermal continuum of the lobes to maintain the structure of the expanding cavity. Here we 
have found direct evidence for thermal particles contributing substantially 
to the internal pressure in the giant lobes of Centaurus~A. Even so, approximate equipartition 
between the thermal and non-thermal gas pressures still cannot entirely explain the 
pressure mis-match in all low-power radio galaxies, indicating that relativistic protons may 
have a substantial contribution in some cases. 

Based on the derived number density of $\sim10^{-4}$~cm$^{-3}$ and the estimated 
volume $V\sim2\times10^{71}$~cm$^{3}$ \citep{hardcastle2009, fermicena}, 
we find that the total mass of the thermal gas within the lobes is 
$M_{th}\sim n_{\rm th} m_H f_V V \sim 2\times10^{10}~M_{\odot}$, 
where $m_H$ is the mass of ionized hydrogen and we use a 
volume filling factor $f_V\sim1$. 
Similar amounts of gas inferred from metal-enriched outflows 
have also been observed in a number of other radio galaxies 
\citep{simionescu2009, kirkpatrick2009, mcnamara2012}. 
We consider the possibility 
that the vast majority of the thermal material has either been entrained 
as the jet ploughed through the interstellar medium (ISM) of the host galaxy 
\citep{laingbridle2002} and/or pushed out by the over-pressured lobes 
as they expand outwards \citep{begelman1989, churazov2001}. 
Entrainment of thermal material is considered as the most likely 
mechanism for decelerating relativistic jets \citep{bicknell1994} 
but the amount of material required is of the order of 
$10^{-3}$~M$_{\odot}$/yr \citep{laingbridle2002}. 
Therefore, this is highly unlikely to be able to provide the 
necessary 200~M$_{\odot}$/yr required by our results, 
for a lobe age of the order of 100~Myr \citep{hardcastle2009}. 
More likely is that the expanding lobes push out large amounts 
of gas from the host galaxy atmosphere/halo through 
successive episodes of jet activity. 
In Centaurus~A there is evidence for at least four episodes of 
activity: the giant outer lobes, the Northern middle lobe \citep{morganti1999}, 
the inner lobes \citep{burns1983}, and the parsec-scale jets \citep{tingay2001}.

Recent simulations \citep{wagnerbicknell2011} have shown 
the importance of inhomogeneity of the ISM which can cause 
deflection of the jet flow, allowing the radio source to effect a 
much larger volume of the host galaxy. In this way, a 
large amount of gas can be swept up by the influence 
of the jet at large distances from the center of the galaxy, possibly 
explaining the substantial amount of gas we have detected within the lobes. 
Such a large removal of gas from the host galaxy provides 
a direct mechanism for suppressing star formation while 
also limiting the amount of gas that can cool and fall 
back towards the center of the galaxy, which may also 
limit the growth of the central supermassive black hole. 
It is also possible that a significant fraction of the mass of thermal gas 
within the lobes was entrained from the intergroup medium 
and dispersed over the entire lobe volume. More detailed simulations 
with synthetic Faraday rotation observations will be required to 
help distinguish between these scenarios. 

\section{Conclusions}
We have presented results from a spectropolarimetric study, using the 
Parkes radio telescope at 1.4~GHz, investigating 
the Faraday rotation of the diffuse polarized emission from the 
giant lobes of the radio galaxy, Centaurus A. Using previous 
results from an RM grid of background radio sources just 
outside the lobes \citep{feain2009}, the smooth 
foreground contribution to the observed RM from Centaurus~A 
was subtracted, leaving a mean residual RM signal of $\sim12$~rad~m$^{-2}$. 

Investigation of whether the residual RM signal comes from a 
thin-skin/boundary layer of magnetoionic material surrounding 
the lobes or from thermal gas internal to the lobes, found the thin-skin 
scenario highly unlikely given that it requires at least an order of 
magnitude enhancement of the swept 
up gas over the expected intragroup density on these scales. 
It should be noted that we cannot conclusively rule out that the residual RM 
signal comes from a degree-scale fluctuation 
in the smooth Galactic foreground that was not sampled by the 
ensemble of background sources outside the lobes and happens 
to align with the orientation of the giant lobes. 

From our investigation of the degree of polarisation at three separate 
frequencies (2.3, 1.4 \& 0.96~GHz) at four positions of high S/N, 
we find strong depolarisation which cannot be explained solely by the 
effects of beam depolarisation. 
Considering all the available data, we consider depolarisation due to 
Faraday rotation within the lobes as our preferred scenario, even in the 
presence of external Faraday dispersion from a foreground screen. 
In this case, we estimate the number density of the thermal gas 
$n_e\sim10^{-4}$~cm$^{-3}$ which gives a total ionized 
gas mass within the lobes of $\sim10^{10}$~M$_\odot$. 
Recent X-ray observations of a small region of the southern lobe also find 
evidence for thermal gas within the lobe with a temperature 
of $\sim0.5$~keV \citep{stawarz2012}. From this we estimate that the thermal 
pressure within the lobes is approximately equal to the non-thermal 
pressure implying that the thermal gas, radio-emitting electrons and 
magnetic field are all in approximate equipartition with each other. 

Future radio polarisation observations are required to have much wider 
$\lambda^2$-coverage to provide the high resolution in Faraday 
depth space needed to uniquely separate the internal and external Faraday 
rotating components.  
Indeed the planned Global Magneto-Ionic Medium Survey (GMIMS) which 
covers 300--1800~MHz \citep{wollebenGMIMS} provides an 
excellent opportunity to resolve the Faraday rotation structure in the 
giant lobes of Centaurus~A. 

\section*{Acknowledgements}

The Parkes radio telescope and Australia Telescope Compact Array 
are funded by the Commonwealth of Australia for operation as 
a National Facility managed by CSIRO.  
B.M.G. acknowledges the 
support of the Australian Research Council through grant FL100100114. 
The authors would like to thank Marijke Haverkorn for her help with 
the observations, as well as Shea Brown, Alex Hill and Leith Godfrey 
for helpful comments and discussions. 
We also thank the referee for providing several important comments 
which helped improve this paper. 
The National Radio Astronomy Observatory is a facility of the National Science 
Foundation operated under cooperative agreement by Associated Universities, Inc.

Facilities:
\facility{Parkes}
\facility{ATCA}


\newpage

\section*{Appendix}

For observations which have more than one source 
of polarized emission along the line of sight with different 
Faraday depths and/or Faraday thicknesses, the measured 
RM can vary with wavelength \citep{law2011, farnsworth2011, osullivan2012}. 
Since we detect both the diffuse polarized emission 
from our Galaxy as well as the polarized emission from Centaurus~A, 
we need to consider how this may affect our measured RM. 
The RM resolution of 310~rad~m$^{-2}$ limits our ability 
to distinguish between multiple regions of polarized emission whose 
Faraday depth differs by less than this amount. 
Therefore, we use models of two line-of-sight 
polarized components with different Faraday depths 
to investigate at what points within 
the lobes of Centaurus~A our measured RMs accurately 
represent the true Faraday depth of the polarized emission 
from Centaurus~A.

To illustrate the effect, we define a two component model 
\begin{equation}
P=p_0\left( \frac{1-e^{-2\sigma_\phi^2 \lambda^4 +2i\lambda^2 \phi}}{2\sigma_\phi^2\lambda^4-2i\lambda^2\phi} \right) e^{2i (\chi_0+\phi_g\lambda^2)} e^{-2\sigma_{\rm RM}^2 \lambda^4} + p_{g} \left( \frac{\sin \phi_{g}\lambda^2}{\phi_{g}\lambda^2} e^{2i(\chi_{0g}+\frac{1}{2}\phi_{g}\lambda^2)} \right) e^{-2\sigma^{2}_{\rm RM}\lambda^4},
\end{equation}
in which the polarized emission of our Galaxy ($p_g$) is described by a 
uniform slab \citep{burn1966} of Faraday depth 
$\phi_g$ while the Centaurus~A model is described in Section~3.5. 
We use the values for position ($13^h27^m$, $-42^{\circ}29'$) listed in Table~2 with 
the Galactic Faraday depth, $\phi_g=-58$~rad~m$^{-2}$ 
taken from the fit for the foreground RM surface. 
The intrinsic polarization angle of the Centaurus~A emission, 
$\chi_0=30^{\circ}$, is taken from a derotated $\lambda=0$ polarization angle at 
this position. For the Galactic emission, we use 
a completely arbitrary value of $\chi_{0g}=90^{\circ}$. 
We then set $p_0=1.0$ and plot the results for various 
fractions of $p_g/p_0$, shown in Figure~7.  

The top panel of Figure~7 shows the Faraday rotation measure 
calculated as ${\rm RM}=d\chi/d\lambda^2$, where 
$\chi=\frac{1}{2}\arctan U/Q$ and $P=Q+iU$. 
This shows that as long as $p_g/p_0 \leq 0.15$ the possible 
systematic error in the measured RM is less than 3~rad~m$^{-2}$,
where the expected RM is $\frac{1}{2}\phi+\phi_g$ 
(solid horizontal line). 
As the Galactic polarized emission (mean value $\sim$52~mJy) 
becomes a larger fraction of the Centaurus~A polarized emission, 
the systematic error in the measured RM becomes much larger (dotted line). 
Since we are unable to disentangle the two contributions 
with our current dataset, we have only included in our calculations 
the regions within the lobes of Centaurus~A where the polarized emission 
is $\gtrsim300$~mJy~beam$^{-1}$. This value roughly corresponds 
to $p_g/p_0 \lesssim 0.15$, which means that even in the regions of 
lowest polarized intensity the potential systematic error in the RM 
is less the mean measurement error within the lobes of $\sim$3~rad~m$^{-2}$.

\begin{figure}[!hb]
\begin{centering}
    \includegraphics[width=8cm]{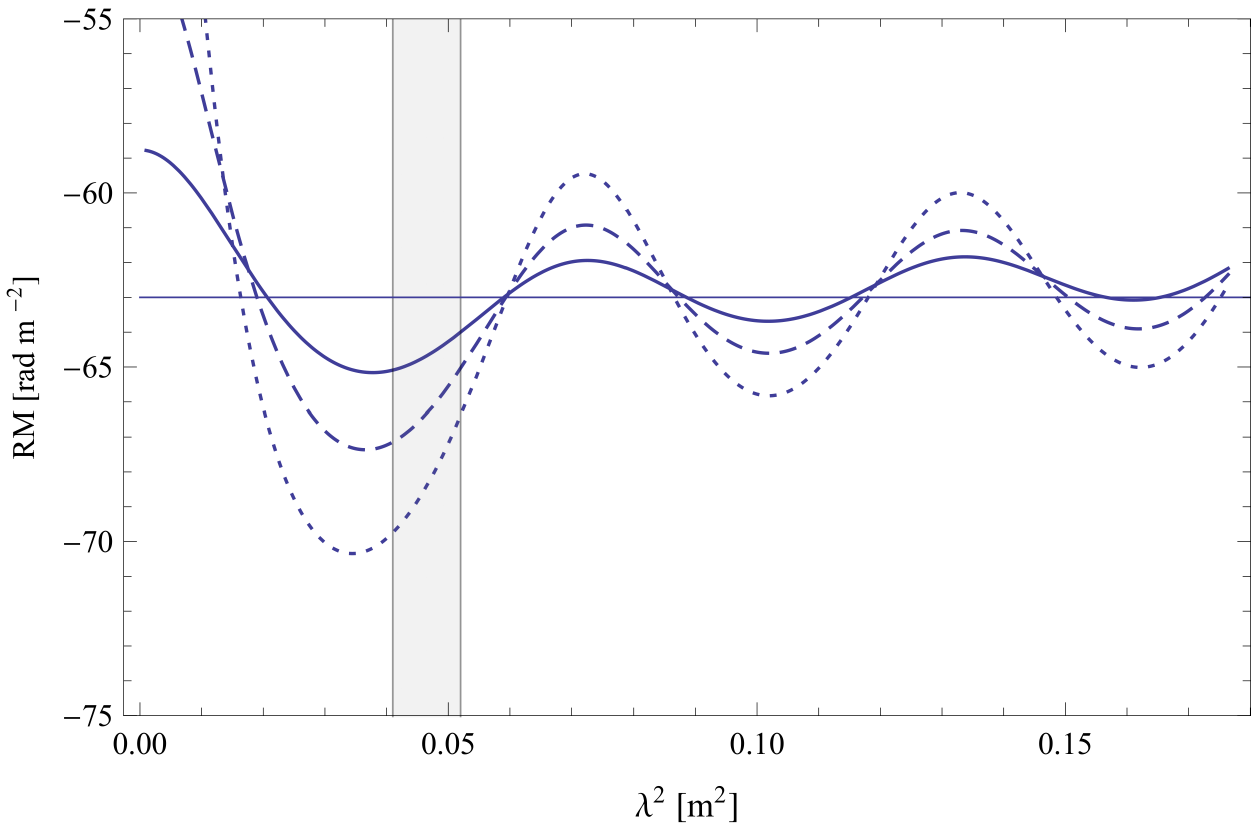}
\caption{Plot of RM vs.~$\lambda^2$, to illustrate the RM variation due to sampling multiple regions of 
  mixed polarized emission and Faraday rotation along a single line of sight. The RM is calculated 
  as $d\chi/d\lambda^2$ and the solid horizontal line is the expected RM (see Appendix for details). 
  Solid line: $p_g/p_c=0.15$, dashed line: $p_g/p_c=0.3$, dotted line: $p_g/p_c=0.5$. 
  The shaded areas indicate the range of $\lambda^2$ space covered by our observations. 
}
\end{centering}
\label{fig_appendixfig_depol}
\end{figure}

\end{document}